\documentclass[acmsmall,nonacm]{acmart}
\usepackage{amsmath,amsfonts}
\usepackage{algorithmic}
\usepackage{graphicx}
\usepackage{textcomp}
\usepackage{xcolor}
\usepackage{listings}
\usepackage{xcolor}
\usepackage{mathtools}
\usepackage{booktabs}
\usepackage{multirow}
\usepackage{listings}
\usepackage[linesnumbered,lined,commentsnumbered]{algorithm2e}
\usepackage{wrapfig}
\usepackage{enumitem}
\usepackage{subcaption}
\usepackage{pifont}
\usepackage{paralist}
\usepackage[compatibility=false]{caption}
\usepackage{multicol}
\usepackage[most]{tcolorbox}
\usepackage{url}
\usepackage[framemethod=TikZ]{mdframed}
\usepackage[commandnameprefix=ifneeded]{changes}

\tcbset{
  myboxstyle/.style={
    colback=gray!20!white, 
    colframe=black,        
    boxrule=0.3mm,         
    arc=0mm,               
    left=1mm,              
    right=1mm,             
    top=1mm,               
    bottom=1mm,            
    fonttitle=\bfseries,   
    coltitle=black,        
    colbacktitle=white     
  }
}

\definecolor{codeRed}{HTML}{AF0000}
\definecolor{codeGreen}{HTML}{008700}
\definecolor{codeOrange}{HTML}{AF5F00}
\definecolor{codeBlue}{HTML}{005FAF}
\definecolor{codePurple}{HTML}{870087}
\definecolor{codeCyan}{HTML}{008787}
\definecolor{codeGray}{HTML}{5F87AF}
\definecolor{black}{HTML}{000000}

\definecolor{IVcol}{HTML}{7030A0}
\definecolor{DVcol}{HTML}{2F5597}
\definecolor{EVALcol}{HTML}{548235}
\definecolor{REFcol}{HTML}{E59500}

\lstdefinelanguage{Spring}
{
    stringstyle=\color{black},
    identifierstyle=\color{codeBlue},
    keywordstyle=\color{codeBlue}\bfseries,
    keywords={server, port, spring, application, name}
}
\lstdefinelanguage{dockerfile}{
    keywords={FROM, RUN, COPY, ADD, ENTRYPOINT, CMD,  ENV, ARG, WORKDIR, EXPOSE, LABEL, USER, VOLUME, STOPSIGNAL, ONBUILD, MAINTAINER},
    keywordstyle=\color{codeBlue}\bfseries,
    identifierstyle=\color{black},
    sensitive=false,
    comment=[l]{\#},
    commentstyle=\color{codePurple}\ttfamily,
    stringstyle=\color{codeRed}\ttfamily,
    morestring=[b]',
    morestring=[b]"
}

\newlist{questions}{enumerate}{2}
\setlist[questions,1]{label=RQ\textsubscript{\arabic*}:,ref=RQ\textsubscript{\arabic*}}

\newcommand{\greybox}[1]{
    \begin{mdframed}[backgroundcolor=black!10!white,linewidth=0pt,backgroundcolor=black!10,linewidth=0pt,innerleftmargin=5pt,innertopmargin=5pt]
        #1
    \end{mdframed}
}
\usepackage{environ}

\newcommand{\cmark}{\ding{51}}
\newcommand{\xmark}{\ding{55}}

\definecolor{configclass}{HTML}{713E5A}
\definecolor{dependencyclass}{HTML}{9CBFA7}
\definecolor{dependencyextract}{HTML}{00798C}

\begin{document}

\title{A Methodology for Evaluating RAG Systems: A Case Study On Configuration Dependency Validation} 
\author{Sebastian Simon}
\affiliation{%
  \institution{Leipzig University}
  \country{Germany}
}
\email{sebastian.simon@cs.uni-leipzig.de}

\author{Alina Mailach}
\affiliation{%
  \institution{Leipzig University, ScaDS.AI Dresden/Leipzig}
  \city{Leipzig}
  \country{Germany}
}
\email{alina.mailach@cs.uni-leipzig.de}

\author{Johannes Dorn}
\affiliation{%
 \institution{Leipzig University}
 \city{Leipzig}
 \country{Germany}
 }
 \email{johannes.dorn@cs.uni-leipzig.de}

\author{Norbert Siegmund}
\affiliation{%
 \institution{Leipzig University, ScaDS.AI Dresden/Leipzig}
 \city{Leipzig}
 \country{Germany}
}
\email{norbert.siegmund@cs.uni-leipzig.de}

\renewcommand{\shortauthors}{Simon et al.}

\begin{abstract}
Retrieval-augmented generation (RAG) is an umbrella of different components, design decisions, and domain-specific adaptations to enhance the capabilities of large language models and counter their limitations regarding hallucination and outdated and missing knowledge. 
Since it is unclear which design decisions lead to a satisfactory performance, developing RAG systems is often experimental and needs to follow a systematic and sound methodology to gain sound and reliable results. However, there is currently no generally accepted methodology for RAG evaluation despite a growing interest in this technology.

In this paper, we propose a first blueprint of a methodology for a
sound and reliable evaluation of RAG systems and demonstrate its applicability on a real-world software engineering research task: the validation of configuration dependencies across software
technologies.

In summary, we make two novel contributions: (i) A novel, reusable methodological design for evaluating RAG systems, including a demonstration that represents a guideline, and (ii) a RAG system, which has been developed following this methodology, that achieves the highest accuracy in the field of dependency validation.
%
%
For the blueprint's demonstration, the key insights are the crucial role of choosing appropriate baselines and metrics, the necessity for systematic RAG refinements derived from qualitative failure analysis, as well as the reporting practices of key design decision to foster replication and evaluation. 

\end{abstract}




\maketitle

\section{Introduction}
Retrieval-augmented generation (RAG) is an emerging novel approach, driven by recent advancements in large language models (LLMs), which have enhanced their capabilities in various software engineering tasks, such as code generation~\cite{chen2021evaluating, xu2022systematic}, configuration validation~\cite{lian2023configuration}, database tuning~\cite{huang2024llmtune, lao2023gptuner}, and program repair~\cite{fan2023automated, joshi2023repair, xia2023automated}. Despite these advancements, LLMs still face challenges, such as hallucination~\cite{zhang2023siren, ji2023survey}, outdated data~\cite{wang2023knowledge, zhang2024comprehensive}, and untraceable reasoning~\cite{kojima2022large, huang2022towards}, which undermine their ability to provide accurate and reliable outputs. RAG offers a promising solution to these challenges by integrating techniques from information retrieval with the generative capabilities of LLMs. This enables the delivery of contextually relevant, up-to-date, and factually accurate information to an LLM, effectively mitigating its limitations.  


Typically, a RAG system comprises of an ingestion part where data is preprocess, embedded, and stored as context documents in a vector database including meta-data, a retrieval part where context documents are retrieved and ranked according to their relevance to a query, and a query part where a prompt with a query is combined with the retrieval results and sent to the LLM. So, RAG is not a single technology, but an umbrella of different components, design decisions, and domain-specific adaptations. It can be composed from different embedding models, search types, filters, and re-rankers. Developers must define which data sources are relevant and how to preprocess them prior to ingestion into a vector database. Moreover, how to prompt the LLM, how to rank which contextual information is most beneficial, and whether and how to provide the LLM with examples to the query at hand, often requires multiple evaluation and refinement steps~\cite{barnett2024seven}. The final RAG system is, thus, a product of several design decisions and iterations based on empirical evidence about the RAG's effectiveness collected on the way. 

Such an iterative, and empirical-driven development process requires a sound research methodology when a RAG system is not only built but also evaluated in a sound way. Without a proper methodology, we may end up in missing opportunities (e.g., due to unconsidered RAG variants), unsound results (e.g., by not being able to causally trace evaluation outcomes to individual RAG design decision), or non-generalizable claims (e.g., by overfitting the RAG adaptions in the iterative development process to the used benchmark). With such a heterogeneity and number of aspects, it can be difficult to not miss important considerations in the study design and evaluation as we will describe in Section~\ref{sec:related}. Although a generally accepted methodology for RAG evaluation clearly calls for a community endeavor, we aim to propose a first blueprint of how such a methodology may look like and apply it on a real-world software engineering research task.

With this paper, we make two novel contributions: (i) a validated blueprint of a reusable empirical study design for evaluating RAG systems in software engineering, and (ii) a RAG system that accurately validates configuration dependencies. 
For (i), the blueprint outlines several empirical key considerations that have to be considered and reported when evaluating RAG, including considerations for independent (context sources, RAG variants) and dependent variables (metrics) variables, as well as aspects of evaluation (baseline, benchmark) and refinement (failure analysis and improvements). This way, we guide researchers and practitioners alike in building a methodological sound study design to evaluate RAG systems. We demonstrate the application and validity of our blueprint through an empirical study on configuration dependency validation.
For (ii), we propose a RAG system (as part of the demonstration) for the specific scenario of dependency validation. Validating configuration dependencies is an important, non-trivial task in practice that requires substantial contextual information due to its project- and technology-dependent variants. Thus, it represents an ideal demonstrator for how to evaluate and conduct experiments with RAG systems and the proposed system is valuable on its own.

These two contributions go hand in hand: It is unsound to propose a blueprint of a methodology without a \emph{real} case of evaluating it. So, rather than using a toy example, we chose an application scenario, where the introduction of a RAG system can make an actual contribution. In other words, the contribution in the area of configuration dependency validation is a significant byproduct of the blueprint's demonstration.
Our reporting on the study design, as well as on the obtained results will therefore take both perspectives. Combined, this serves also as a guide for using the proposed blueprint.

The main result is that by following the blueprint, we were able to systematically develop and evaluate a RAG system that achieves the highest accuracy in the field of configuration dependency validation (e.g., compared to~\cite{lian2023configuration} on a similar task).
A key insight of our blueprint is the importance of selecting the right baseline and metrics to determine whether the introduction of RAG is truly beneficial for a given task, because we found in our study that baselines can surpass RAG systems. Moreover, it emphasizes the need for systematic refinements (e.g., derived from failure analysis) to unlock RAG's full potential as we found significant improvements when refining the RAG based on a qualitative failure analysis. Finally, the blueprint ensures that key design decisions are reported, facilitating better replication, evaluation, and identification of best practices and common failure points in RAG systems.


\newpage

In summary, we make the following contributions:
\begin{compactitem}
    \item A blueprint for a sound empirical study design to evaluate RAG systems; 
     \item A demonstration of how this study design can be applied;
    \item A highly accurate RAG system for configuration dependency validation.
    \item A replication package containing all code scripts, datasets, prompt templates, and validation results (see \hyperref[sec:data-availability]{Data Availability} section for details).
\end{compactitem}

\section{Scenario and Related Work}
This section introduces configuration dependency validation, reviews existing approaches, and identifies real-world challenges where a RAG system could be beneficial. We then present related work on RAG for software engineering tasks, RAG system evaluation, and prompt engineering.

\subsection{Scenario: Configuration Dependency Validation}
Modern software development requires coordinating a wide range of technologies, such as code frameworks, build tools, and databases~\cite{sayagh2017cross}. This coordination usually revolves around configuring each technology in its own format. For instance, container configurations are defined in a \texttt{Dockerfile}, whereas Java application options are set in a \texttt{applications.yml}. These configuration files often encode hundreds of configuration options in their own structure, syntax, and semantic~\cite{siegmund2020dimensions}. Most of these diverse technologies' configurations need to be harmonized to ensure their correct interplay and interoperability. In the exemplary Spring Boot application scenario shown in \autoref{fig:configExample}, the port configurations in the \texttt{application.yml} in Listing~\ref{lst:example-application} and in the \texttt{Dockerfile} in Listing \ref{lst:example-docker} must match for the application to work. This constraint constitutes a \textit{configuration dependency}. Such dependencies can exist both within a single technology (\emph{intra-}technology) and across multiple technologies (\emph{cross-}technology).

\begin{figure}[ht]
\vspace{-0.5\baselineskip}
\centering
 \begin{minipage}[t]{0.48\columnwidth}
  \centering
  \begin{lstlisting}[language=Spring, label=lst:example-application, caption=Spring Boot's application.yml specifying the port of the web server.]
server:
  port: 8080
spring:
  application:
    name: app
\end{lstlisting}
 \end{minipage} \hfill 
 \begin{minipage}[t]{0.48\columnwidth}
   \centering
    \begin{lstlisting}[language=Dockerfile, label=lst:example-docker, caption=A Dockerfile exposing the port of the web server.]
FROM openjdk:11-jdk-slim
ARG JAR_FILE=target/*.jar
ADD ${JAR_FILE} app.jar
EXPOSE 8080
CMD ["java", "-jar", "/app.jar"]
\end{lstlisting}
 \end{minipage}
 
\vspace{-0.25\baselineskip}
\caption{An exemplary cross-technology configuration dependency between Spring Boot and Docker, both specifying the port of the Web server.}
\label{fig:configExample}
\end{figure}

\noindent Naturally, developers of these technologies cannot meticulously manage all the frameworks and tools that potentially interact in their technology stack. Consequently, configuration dependencies are rarely fully documented~\cite{chen2020understanding} and the corresponding documentations are commonly incomplete and outdated~\cite{hubaux2012user, jin2014configurations, rabkin2011static}. This lack of a comprehensive overview about how and where configuration dependencies manifest within a software project poses several challenges in practice: (1)~the ever-evolving configuration landscape of software projects bears the risk of introducing severe misconfigurations due to the violation of dependencies ~\cite{maurer2015fail, tang2015holistic, mehta2020rex}; (2)~misconfigurations may go unnoticed into production due to the interplay of multiple technologies that manifest themselves only in combination; (3)~resulting configuration errors are complex and far-reaching since several technologies and diverse configuration artifacts are involved~\cite{yin2011empirical, xu2015systems}; (4)~detecting and resolving these misconfigurations is often more challenging than fixing software bugs, making misconfigurations one of the most common causes of software failures in production today~\cite{oppenheimer2003internet, gunawi2016does, rabkin2012hadoop}. 


Several dependency-detection approaches have been developed, primarily leveraging heuristics~\cite{simon2023cfgnet, chen2020understanding, chen2016determine, xu2013not}, domain-specific languages~\cite{huang2015confvalley, baset2017usable}, or machine learning techniques~\cite{santolucito2017synthesizing, santolucito2016probabilistic, zhang2014encore}. However, the detection process of configuration dependencies is prone to producing many false positives by incorrectly linking configuration options that are identical only coincidentally, but, in fact, independent from one another. This significantly impairs the practicality of dependency-detection approaches, because developers need to be able to trust the accuracy of warnings and errors produced by tools to support them.
The goal of our proposed RAG system is to overcome the challenges of false positives by providing accurate classification capabilities in a technology-agnostic way. 
So, configuration validation represents a prime example of a complex software engineering task that could be improved by RAG and we will use it to demonstrate how RAG systems can be systematically evaluated.

\subsection{Related Work}\label{sec:related}
Next, we discuss existing approaches that use a RAG system for software engineering task and that are concerned with the evaluation of RAG systems, including prompt engineering techniques.

\paragraph{RAG for Software Engineering Tasks}
RAG is on the brink to become a mayor technique for solving software engineering tasks. This is reflected in the high number of papers and pre-prints that are just being released. Although partially not peer reviewed,  at least, we can derive what topics are addressed and for which use cases RAG provides a potential improvement.
For instance, despite the wide adoption of LLM-based tools to generate code~\cite{peng2023impact}, they frequently hallucinate APIs that do not exist and lead to errors~\cite{liu2024exploring}. \citeauthor{jain2024mitigating}~\cite{jain2024mitigating} substantially reduce hallucination by providing the LLM with additional documentation context.
Improving LLM generations using retrieved context has also been proposed to improve commit message generation~\cite{shi-etal-2022-race, xue2024automated, zhang2024rag}, test case generation~\cite{mackay2024test}, and to establish traceability between natural language requirements and code artifacts~\cite{aliestablishing}. 

Although these studies present advances achieved by leveraging RAG, they do not follow a standard methodology or terminology: Some vary different components of the RAG system and some vary the LLM that is used for generation. Similarly, a wide range of metrics is used to measure effectiveness, ranging from task-specific metrics to general NLP metrics.
Many papers omit details on iterative system refinement. Some mention experimental development phases (e.g., prompt improvement~\cite{aliestablishing} or RAG architecture configuration~\cite{mackay2024test}) while others fail to report systematic approaches or measures to ensure result validity.
As there are numerous design decisions within a RAG, it is hard to decide upfront what factors to vary and so they become naturally part of the study. However, when this is not systematically done and properly reported, validity and replicability is threatened.
Our work aims at supporting researchers with this decision process by providing a blueprint of a methodology on how to conduct a sound empirical evaluation of a RAG system.

\paragraph{Evaluating RAG}
Evaluating RAG systems, particularly their retrieval and generation components, has been addressed in several recent studies. For example, \citeauthor{yu2024evaluation}~\cite{yu2024evaluation} provide a comprehensive overview about the evaluation and benchmarks of RAG systems, highlighting challenges in evaluating RAG systems and presenting existing quantifiable metrics of the retrieval and generation components.
\citeauthor{es2023ragas}~\cite{es2023ragas} propose RAGAS, a framework introducing a suite of metrics to evaluate the different dimensions in a RAG pipeline, with an emphasis on retrieval and generation metrics, such as context or answer relevance. 
\citeauthor{salemi2024evaluating}~\cite{salemi2024evaluating} focus on evaluating retrievers in RAG systems. They propose eRAG, an evaluation approach for retrievers, where the LLM in the RAG system is applied on each document in the retrieval list and set-based or ranking metrics are used as aggregation function to obtain an evaluation score for each retrieval list. Their approach significantly offers computational advantages compared to end-to-end evaluations. Furthermore, \citeauthor{chen2024benchmarking}~\cite{chen2024benchmarking} conduct a comprehensive evaluation of RAG for current LLMs. They introduce Retrieval-Augmented Generation Benchmark (RGB), a new corpus for RAG evaluation on abilities of LLMs, such as noise robustness, negative rejection, information integration, and counterfactual reasoning. While all these approaches provide valuable solutions to the evaluation of RAG systems, particularly their retrieval and generation components, they do not help researchers with designing sound experiments. That is, they are concerned with the RAG components rather than how to address a research question in a sound way. We propose a blueprint of a methodology on how to conduct a sound empirical evaluation of RAG systems, supporting developers and researchers alike making sound and reasonable decisions. 


\paragraph{Prompt Engineering}
Prompts are specific instructions provided to an LLM to generate desired responses from it. Empirical investigations have shown that providing different prompts to an LLM has a significant impact on its performance~\cite{kojima2205large, zhou2022large}. That is, ad-hoc developed prompts do not consistently lead to desired results, while carefully crafted prompts can unleash the vast capabilities of an LLM to deliver remarkable results. Systematically crafting and designing the best prompt to accomplish a given task is known as \emph{prompt engineering}~\cite{reynolds2021prompt}.
The probably most popular and widely used prompting strategies are zero-shot~\cite{radford2019language}, few-shot~\cite{brown2020language} and chain-of-thought prompting~\cite{wei2022chain}. Zero-shot prompting relies on carefully crafted prompts that guide an LLM towards a specific task while few-shot prompting involves the introduction of few input-output example of the given task to demonstrate the LLM how to solve a task.
Prompt engineering is thus also a crucial part of a RAG system to improve its output quality~\cite{zhao2024retrieval}. 

\section{A reusable Methodology for RAG evaluation}

In this section, we present a blueprint of a sound empirical study design for evaluating RAG systems. Following, we describe the key considerations of our blueprint, which have to be discussed and reported when evaluating RAG.
We not only propose the blueprint, but also provide a concrete demonstration on how to implement and report decisions in the remainder of the paper.


\begin{figure}[tbh]
    \centering
    \includegraphics[width=0.999\linewidth, trim=0.05cm 0.25cm 0.1cm 0.12cm, clip]{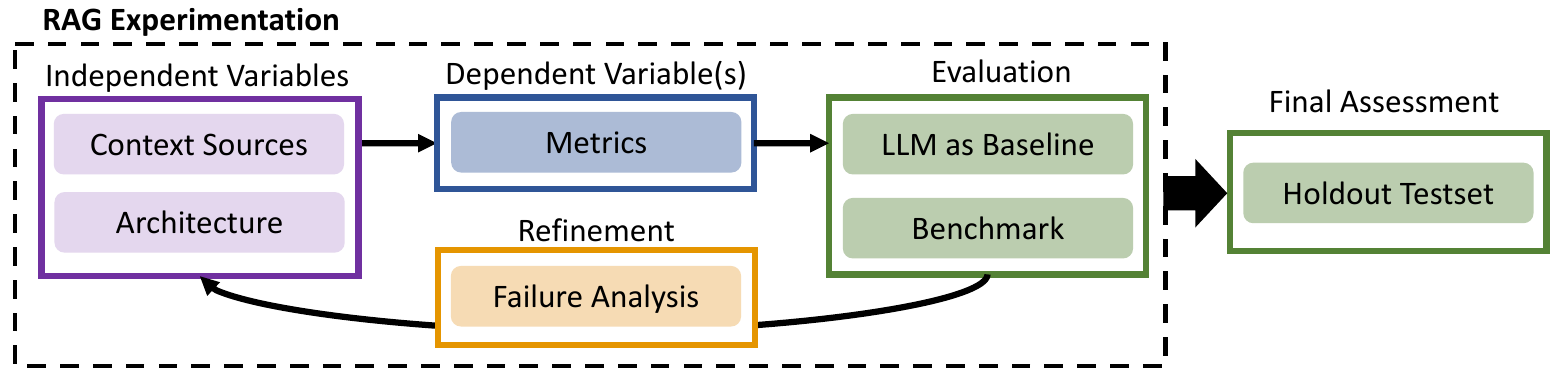}
    \caption{Key considerations for sound empirical evaluations and refinements of RAG systems}
    \label{fig:blueprint}
\end{figure}

Figure~\ref{fig:blueprint} highlights the key aspects to consider when conducting an empirical evaluation. We use the empirical terminology of dependent and independent variables to define the factors influencing the evaluation process and outcomes. For each consideration, we will now provide guiding questions and discussions.

\subsection*{Independent variables: What do we vary?}

\paragraph{\textcolor{IVcol}{Context resources}} Providing different context resources is the key difference between a RAG system and a vanilla LLM as they represent the factual and timely data on which an LLM can reason on instead of only using pretrained weights based on the training data. It is, thus, crucial to select resources that are relevant to the task at hand, which usually requires human judgment. Moreover, as different resources, such as Web searches, database entries, user feedback, and historical data, all can contribute differently to the output accuracy, it is important to evaluate their individual and combined contribution to the independent variables.

\paragraph{\textcolor{IVcol}{Architecture}} RAG is an umbrella of different components, design decisions, and domain-specific adaption. For example, there are hundreds of different embedding models to encode the semantics of textual information for later search~\cite{wang2024searching}, there are multiple ways for searching relevant documents in a multitude of vector stores~\cite{huang2024survey}, and there are plenty of ways on how to preprocess information (e.g., splitting documents), as well as how to filter and re-rank retrieved documents~\cite{gao2023retrieval}. All these \emph{architectural} decisions in a RAG system can influence result accuracy and therefore need to be considered in a sound empirical study. As discussed in Section~\ref{sec:related}, there is a current lack in reporting on design decisions, as well as improvements and refinements on them. Whether a proposed RAG system is a first, naive version or a carefully crafted one based on multiple iterations and feedback runs remains unclear, but is important on estimating the (un)tapped potential of the RAG system.

\subsection*{Dependent variables: What and how to measure?}
\paragraph{\textcolor{DVcol}{Metrics}} Every study begins with a research question, which then must be operationalized. If the goal is to assess the \emph{effectiveness} of a RAG system for a specific task, this is often translated into task-specific metrics. For classification, metrics, such as precision or recall may be suitable, whereas generation tasks typically lack straightforward measures. RAG-specific metrics, such as context relevancy, assess how relevant retrieved documents are to the generation task. Although more relevant documents should improve RAG effectiveness, it is essential to empirically verify that a metric correlates with the desired outcome. So, RAG-specific metrics may be a part of the experiment, but due to their independence to the task, are usually insufficient to answer the research questions. Instead, defining metrics that truly reflect the system's performance on the intended task should be the main priority. 

\subsection*{Evaluation: How to evaluate the RAG system?}
 \paragraph{\textcolor{EVALcol}{Baseline}} A \emph{baseline} provides a point of reference to determine whether the introduction of a RAG system is beneficial at all. Research has shown that simple machine learning baselines, when handled correctly, often outperform more complex models~\cite{fu2017easy}. While creating a baseline can sometimes be challenging, it is crucial to ensure that system improvements genuinely enhance effectiveness and that unnecessary complexity is avoided when simpler approaches yield similar results. For RAG systems, using a vanilla LLM as a baseline is often appropriate, but the baseline must be refined alongside the RAG system for a fair comparison (e.g., if context-independent prompting of the RAG system is improved, the baseline prompting has to be improved as well). An analogy is hyperparameter tuning. When tuning only the proposed approach, but not the baseline, it remains unclear whether differences are due to the tuning or the approach itself. In cases where retrieval is essential, a basic RAG version can serve as the baseline instead of a vanilla LLM. One final aspect is the need to report the version number of an LLM. When only the architecture, such as GPT-4 or GPT-3.5 is reported, but not the version, we will not be able to reproduce the results as updates on the models change substantially their behavior and quality.

\paragraph{\textcolor{EVALcol}{Benchmark}} To compute the metrics for the dependent variable, both the RAG system and the baseline must be evaluated on a \emph{benchmark}. A benchmark provides a standardized basis for comparing different approaches. For well-researched use cases, established benchmarks may already exist. In other cases, it may be necessary to create or adapt a benchmark. To allow future researchers to compare their approaches to the presented one, a benchmark should be well-documented and publicly available. This ensures transparency and facilitates reproducibility in future work. Of course, this is not always possible. So, at least, the evaluation data set should be made public.


\subsection*{Refinements: How to systematically refine the RAG system?}

\paragraph{\textcolor{REFcol}{Failure Analysis}} 
A RAG system may benefit from several iterations on the task evaluation, in which each iteration may come with a \emph{refinement} of all RAG components, their data sources, and used prompt templates. 
To gain meaningful insights into how to refine the system and identify the most promising changes, analyzing failure cases is essential. Depending on the use case, errors may require different analytical approaches. For classification problems, performance across classes can provide valuable insights, but in most cases, manual qualitative inspection will be necessary for deeper understanding. This analysis can be guided by questions such as: Is the context properly utilized? What additional information would the LLM need to solve certain cases successfully? Are the tasks clearly specified for the LLM? Such a failure analysis should be properly reported in the study to enable traceability of RAG decisions. As said before, it is important to apply changes (if possible) to the vanilla LLM as well. 

\subsection*{Final Assessment: How to ensure validity and soundness after refinement and re-evaluation?}

\paragraph{\textcolor{EVALcol}{Holdout Testset}} Each refinement requires an evaluation again. The benchmark or evaluation set from the first evaluation are likely candidates, but since the failure analysis has been conducted on the same data, we risk overfitting the system to this dataset. Then it becomes unclear whether observed improvements generalize to new data, essentially limiting generalizability and threatening external validity. Maintaining a separate \emph{holdout testset} serves as an unseen, labeled dataset that helps to assess the effect of the \emph{final} RAG system. As soon as this holdout testset is used and a refinement follows, we have to create a new testset. The final reporting about the RAG's performance should be mainly focused on the results of this testset. 

\section{Demonstration of Evaluating RAG for Configuration Dependency Validation}
In this section, we introduce an approach to leverage RAG for configuration validation and demonstrate how our reusable research methodology can be applied to evaluate this approach.

\subsection{Configuration Dependency Validation with RAG}
We propose a novel RAG system to validate configuration dependencies. In the following section, we describe its architecture from indexing data as context into a vector database to retrieving relevant context from it, and querying an LLM with an augmented prompt.

\paragraph{Context Ingestion}
We populate a vector database with \emph{static} and \emph{dynamic} context information. \emph{Static} context information is stored prior to the evaluation queries. \emph{Dynamic} context information is dynamically extracted, processed, and indexed into the vector database on-the-fly for each dependency during validation.

The indexing process starts with transforming the content of \emph{static} and \emph{dynamic} context sources into documents, which are generic containers that stores text along with metadata. Technically, we use a data loader from llamaindex\footnote{\url{https://www.llamaindex.ai/}}. The resulting documents are then split into smaller chunks with a certain amount of overlap. We choose a chunk size of 512 with an overlap of 50 in order to encompass all necessary information in the retrieved chunks. After splitting, the resulting chunks are then converted into embeddings, a vector representation of the documents, using an embedding model. Finally, the resulting embeddings are indexed in the vector database Pinecone\footnote{\url{https://www.pinecone.io/}}.

\paragraph{Context Retrieval}
We incorporate several retrieval strategies to enhance retrieval quality of our RAG system. First, we optimize the original query by rewriting. We then perform a hybrid search in the vector database with the retrieval query to obtain an initial set of contextual information. Hybrid search combines semantic search with keyword search in one query, which allows us to find context information based on configuration option names via keyword search and relevant context of dependencies via semantic search. After obtaining a large initial set of context information, we re-rank the elements of this set according to their relevance. Finally, we use the re-ranked contextual information to augment the prompt for dependency validation.

\begin{figure}
    \centering
    \includegraphics[width=\linewidth, trim=0.00cm 0.25cm 0.15cm 0.12cm, clip]{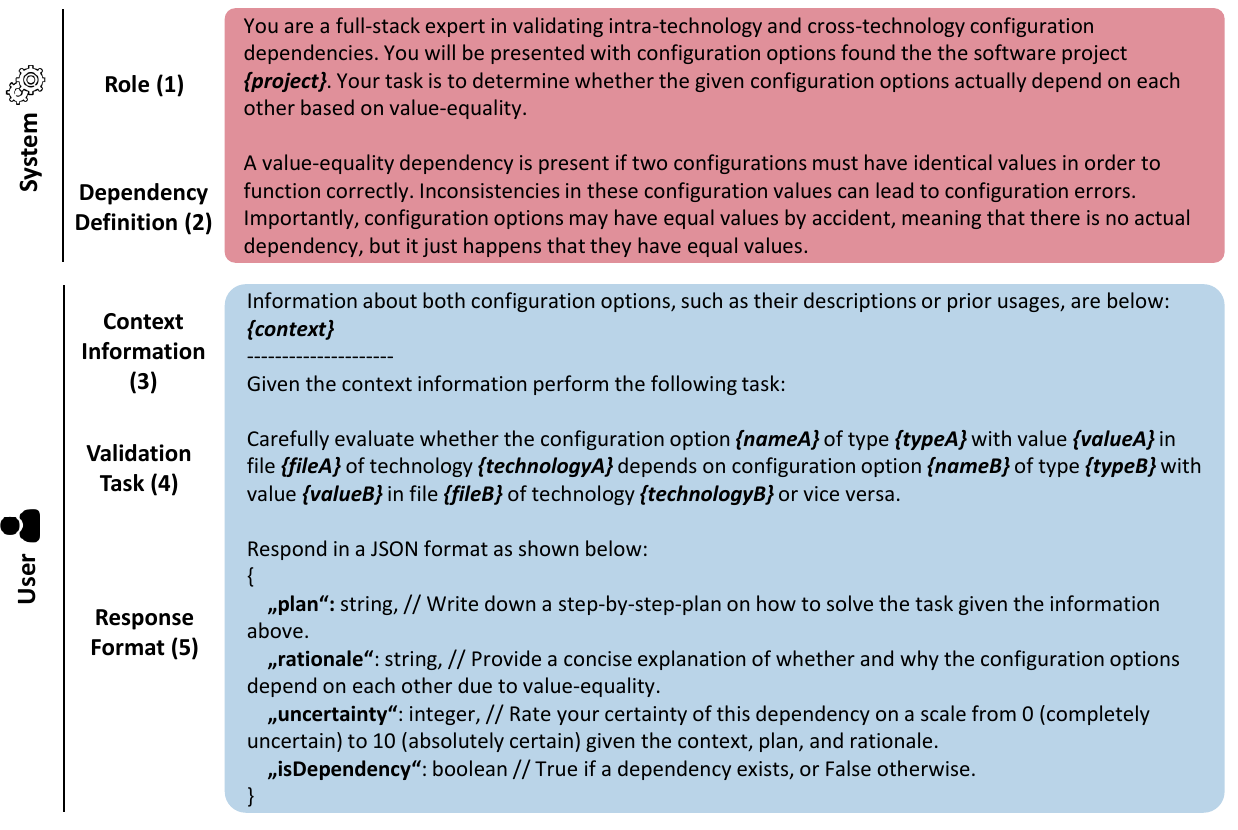}
    \caption{The prompt template and its components used for dependency validation.}
    \label{fig:prompt}
\end{figure}

\paragraph{Validation Generation}
We created a prompt for dependency validation using a collective prompt design process, in which all authors discussed and adjusted the prompt and its components in multiple iterations. Our final prompt for dependency validation includes five elements: (1) a system message, which defines the role of the LLM, (2) a definition of the dependency that it has to validate, (3) the retrieved context information, (4) the actual instruction to validate the given dependency, and (5) a description of the JSON format in which the LLM should respond. The validation prompt and its components are shown in \autoref{fig:prompt}.

The JSON response format prompts the LLM to fill in four primary fields:
(1) \emph{plan}: a string that describes the plan on how to validate the dependency based on the contextual information step-by-step; (2) \emph{rationale}: a string that explains whether and why the configuration options depend on each other or not; (3) \emph{uncertainty}: an integer from 0 (completely uncertain) to 10 (absolutely certain) indicating the LLM's uncertainty about the dependency based on the context information; and (4) \emph{isDependency}: a boolean indicating the final decision of the LLM whether the given dependency is a true dependency or a false positive. 

The proposed response structure follows the idea of chain-of-thought prompting~\cite{wei2022chain}, aiming to increase the understanding and analytical capabilities of LLMs in validating configuration dependencies. That is, we prompt the LLM to lay out a step-by-step strategy on how to validate the given dependency and to provide an explanation and rating of its uncertainty of before making the final decision. By breaking down the validation task into several steps, each building on the previous one, we ensure a logical progression of actions that the LLM should perform. 

\subsection{Research Questions}
To evaluate the effectiveness of the proposed RAG system and identify key design decisions that enhance dependency validation, we conduct an experimental study. This study aims to address the following two research questions:

\begin{compactitem}
\setlength{\itemsep}{0em}
\item
\textbf{RQ\textsubscript{1}}:
How effective are vanilla LLMs and unrefined RAG in validating configuration dependencies?
\item
\textbf{RQ\textsubscript{2}}:
What validation failures occur with vanilla LLMs and unrefined RAG and what refinements can be derived from that?
\end{compactitem}
By asking and answering these questions, we contribute to advancements in configuration dependency validation while also completing a full cycle of a RAG evaluation and refinement, demonstrating the applicability of our proposed blueprint.

\subsection{Applying the Blueprint: Evaluating RAG for Configuration Dependency Validation}

\begin{figure}[th]
    \centering
    \includegraphics[width=\linewidth, trim=0.1cm 0.16cm 0.16cm 0.2cm, clip]{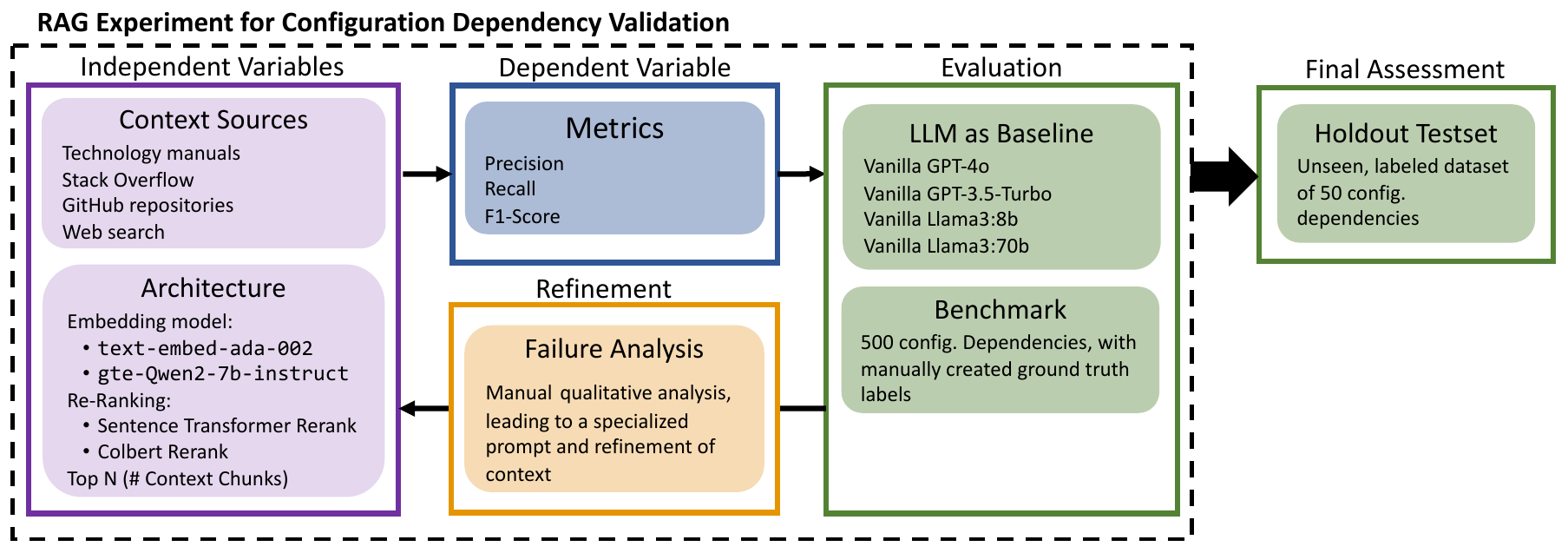}
    \caption{Study design decisions based on the blueprint for evaluating RAG for configuration dependency validation.}
    \label{fig:blueprint_instance}
\end{figure}

By using our blueprint, we can give a concise overview in Figure~\ref{fig:blueprint_instance} of how we addressed several design choices for conducting the evaluation. We will now discuss each decision in detail. 

\subsubsection{Independent variables}

\paragraph{\textcolor{IVcol}{Context Sources}} Data on configuration dependencies is rarely available and usually not explicitly stated, which makes deciding on context sources especially challenging. For example, manuals typically describe configuration options in terms of their names, abbreviations, functionality, and (rarely) their allowed parameter ranges. But the description almost always lacks explicit dependency information to other options of that technology and entirely to other technologies. Hence, there is no clear source for reliable information on configuration dependencies. Developers often turn to the man pages, Stack Overflow, or Web searches that provide real-world use cases of configuration options in which concrete dependencies are explicitly or implicitly discussed. Furthermore, project-specific information, such as project and environment settings in which the configuration resides, may be found in the corresponding project's repository. Based on these considerations, we decide to collect information from the same sources that also developers use, that is, man pages, Stack Overflow posts, GitHub repositories, and Web search results. 

\begin{wraptable}[10]{r}{5.0cm}
\vspace{-1.0\baselineskip}

\footnotesize
\caption{Variants of RAG used in the first experiment with text-embed-ada-002~(ada2) and gte-Qwen2-7B-instruct~(Qwen2).}
\label{tab:rag-configs}
\renewcommand{\tabcolsep}{2.25pt}
\begin{tabular}{@{}clrlr@{}}
\toprule
ID & Embed. & Embed. & Reranking & Top \\
   & Model     & Dim.   &    & N   \\ \midrule
1  & ada2    & 1536 & Colbert              & 5  \\
2  & Qwen2 & 3584 & Colbert              & 5  \\
3  & Qwen2 & 3584 & Sentence Transf. & 5  \\
4  & Qwen2 & 3584 & Colbert              & 3  \\ \bottomrule
\end{tabular}
\vspace{-1.0\baselineskip}
\end{wraptable}

\paragraph{\textcolor{IVcol}{RAG Architecture}}
We vary between several core elements, such as the embedding model (ada2 vs. Qwen2) , the re-ranking algorithm (Colbert vs. Sentence Transformer), and the final number of context chunks provided to the LLM (3 vs. 5). By varying only one component at a time, we obtain four different RAG variants, shown in Table~\ref{tab:rag-configs}. Note, that we clearly could manipulate many more aspects of the RAG pipeline. However, this is the first and unrefined setup with which we start the evaluation, these configurations might be refined during experimentation if the failure analysis reveals significant room for improvement.

\subsubsection{Dependent Variable}

\paragraph{\textcolor{DVcol}{Metrics}} The dependent variable in our demonstration is the effectiveness of the approach in validating configuration dependencies. We measure effectiveness using traditional classification metrics: precision~(\ref{eq:precision}), recall~(\ref{eq:recall}), and F1-score~(\ref{eq:f1}). Precision represents the proportion of true dependencies (TP) among positively labeled ones (TP + FP), while recall reflects the proportion of true dependencies among all actual dependencies (TP + FN). These metrics highlight the trade-off between identifying as many true dependencies as possible (recall) and ensuring the correctness of positively labeled dependencies (precision). High precision and recall are crucial for trustworthy validation. The F1-score, the harmonic mean of precision and recall, provides a single scalar value while allowing for more detailed analysis when examining each metric individually.
\begin{subequations}
\begin{minipage}{0.275\linewidth}
\begin{equation}
    \text{Precision} = \frac{\text{TP}}{\text{TP}+\text{FP}}\label{eq:precision}
\end{equation}
\end{minipage}
\hfill
\begin{minipage}{0.245\linewidth}
\begin{equation}
    \text{Recall} = \frac{\text{TP}}{\text{TP}+\text{FN}}\label{eq:recall}
\end{equation}
\end{minipage}
\hfill
\begin{minipage}{0.345\linewidth}
\begin{equation}
    \text{F1} = \frac{2 \cdot (\text{Precision} \cdot \text{Recall})}{(\text{Precision}+\text{Recall})}\label{eq:f1}
\end{equation}
\end{minipage}
\end{subequations}
\begin{samepage}
\subsubsection{Evaluation}
\paragraph{\textcolor{EVALcol}{Baseline}}
\end{samepage}
To assess the effectiveness of our RAG variants for dependency validation, we compare their validation abilities against vanilla LLMs. Throughout our empirical study, we leverage four state-of-the-art LLMs, including two proprietary LLMs developed by OpenAI\footnote{\url{https://openai.com/}} (i.e., GPT-4o and GPT-3.5-Turbo) and two open-source LLMs developed by Meta\footnote{\url{https://llama.meta.com/llama3/}} (i.e., LLama3:8b and LLama3:70b (non-quantized)). The open-source LLMs are executed on our own hardware, a MacBook Pro with an Apple M3 chip and 128GB of RAM. A summary of the studied LLMs with their properties can be found in Table~\ref{tab:models}. For all LLMs, we set the temperature to 0 in order to improve determinism and consistency of LLM outputs. Note that the LLMs play a second role here. They are not only baselines, but also independent variables as we vary and compare them (to each other).

\begin{wraptable}{r}{5.5cm}
\centering
\footnotesize
\caption{Overview of studied LLMs}
\label{tab:models}
\renewcommand{\tabcolsep}{2pt}
\begin{tabular}{@{}lrrc@{}}
\toprule
Model  & \multicolumn{1}{c}{\#\,Param} & \multicolumn{1}{c}{Context} & Open \\
Name       & \multicolumn{1}{c}{(B) }     & \multicolumn{1}{c}{Length}  & Source \\ \midrule
gpt-4o-2024-05-13     & -        & 128k    & \xmark \\
gpt-3.5-turbo-0125    & 175      & 16k     & \xmark \\
llama3:70b & 70       & 8k      & \cmark \\
llama3:8b  & 8        & 8k      & \cmark \\ \bottomrule
\end{tabular}
\vspace{-1.0\baselineskip}
\end{wraptable}
\paragraph{\textcolor{EVALcol}{Benchmark}}
To evaluate the effectiveness of our approach in validating configuration dependencies, we collected a dataset of potential configuration dependencies. We ran \emph{CfgNet}, an existing dependency detection framework from the literature~\cite{simon2023cfgnet}, on ten real-world software projects. The selected projects used at least the following technologies: Spring Boot, Maven, Docker, and Docker Compose, and were among the most starred on GitHub. An overview of the systems is provided in Table~\ref{tab:subject-systems}. 
For each software project, we sampled 50 potential dependencies, leading to a final set of 500 candidates. The rational for this sampling is that all dependencies must be manually checked in detail to create a trustworthy and correct ground truth. We manually labeled each dependency as either \emph{true} if it correctly represents a dependency, or \emph{false} if it is a false positive. This way, we created the largest ground truth cross-technology configuration dependency dataset that we are aware of and that has been used in any study. As a notable contribution of this work, we make this dataset available for other researchers and tool developers.


\begin{wraptable}[15]{r}{6.2cm}
\vspace{0.750\baselineskip}
\footnotesize
\caption{Subject Systems with Version (SHA), Number of Stars ($\star$), Number of Commits (Com.)}
\label{tab:subject-systems}
\renewcommand{\tabcolsep}{2pt}
\begin{tabular}{@{}lrrr@{}}
\toprule
GitHub Project                & Version & $\star$ & Com. \\ \midrule
macrozheng/mall               & 70a226f & 76.6k & 1k \\
apolloconfig/apollo           & 49bd8cc & 28.9k & 2.8k \\
linlinjava/litemall           & 92ffc39 & 19k & 1.2k \\
sqshq/piggymetrics            & 6bb2cf9 & 13.1k & 290 \\
codecentric/spring-boot-admin & 60be5d1 & 12.2k & 3.1k \\
macrozheng/mall-swarm         & fd5246c & 11.5k & 463 \\
wxiaoqi/Sping-Cloud-Platform  & 9aad435 & 6.3k & 358k \\
pig-mesh/pig                  & 8e10249 & 5.7k & 1.6k \\
jetlinks/jetlinks-community   & 1ad1e44 & 5.2k & 1.2k \\
Yin-Hongwei/music-website     & 12e1b0a & 5.1k & 385k \\ \bottomrule
\end{tabular}
\vspace{-1.0\baselineskip}

\end{wraptable}

\subsubsection{RAG Refinement}

\paragraph{\textcolor{REFcol}{Failure Analysis}}
We conducted a qualitative analysis of the validation failures of vanilla LLMs and the best-performing RAG variant to identify failure patterns. Specifically, we inspected all validation failures by extracting the involved configuration options, the retrieved context, and the reasoning of the LLM. For example, we reviewed the options' names, their values, and the involved technologies. Then, we checked whether the reasoning of the LLM was consistent with its final (wrong) assessment. We put this reasoning with our own manual correct classification in relation to find, for instance, whether additional information was not available to the LLM to make a correct judgment or whether false information, noise, or ambiguities caused the failure. We then categorize failures by similar mistakes or causes of mistakes.The following refinement step used these categorizes to implement measures for avoiding the failure causes.


\subsubsection{Final Assessment}

\paragraph{\textcolor{EVALcol}{Holdout Testset}} In addition to the benchmark consisting of 500 dependency candidates, we created a holdout testset of 50 potential dependencies by randomly sampling configuration dependencies from our subject systems and manually reviewing them similar to how we labeled the dependencies of the baseline. The final assessment of whether our proposed RAG system is able to accurately validate configuration dependencies is conducted and reported on this testset.

\section{Results}

This section presents the results of our empirical study, which has two goals: First, we demonstrate the applicability and validity of our RAG methodology. Second, we evaluate the effectiveness of RAG and vanilla LLMs on the specific scenario of dependency validation, as an independent contribution on its own. Results are structured by the scenario's research questions. We discuss the implications and insights for our proposed methodology alongside. 

\subsection{RQ1: Validation Effectiveness of Vanilla LLMs and unrefined RAG on Dependency Validation}
%


\begin{wraptable}[32]{r}{0.44\textwidth}
\vspace{-1.\baselineskip}
\caption{Validation effectiveness of vanilla LLMs GPT-4o (4o), GPT-3.5-Turbo (3.5T), Llama3:8b (L3:8b), Llama3:70b (L3:70b), and different RAG variants on 500 configuration dependencies extracted from ten real-world software projects.}
\footnotesize
\label{tab:validation-effectiveness-all}
\renewcommand{\tabcolsep}{2.2pt}

\begin{tabular}{@{}p{0.5cm}lrrrrr@{}}
\toprule
RAG ID & LLM & \#Failures & \multicolumn{1}{l}{Precision} & \multicolumn{1}{l}{Recall} & \multicolumn{1}{l}{F1-Score} \\ \midrule
\multirow{5}{*}{w/o}       & 4o                 & 91  & 0.89                        & 0.62                       & 0.73                         \\
                           & 3.5T               & 157 & 0.59                        & 0.67                       & 0.63                         \\
                           & L3:70b             & 154 & 0.63                        & 0.56                       & 0.59                         \\
                           & L3:8b              & 159 & 0.55                        & 0.84                       & 0.65                         \\
                           \noalign{\vskip -\aboverulesep}\cmidrule{2-6}\noalign{\vskip -\belowrulesep} 
                           & mean               & 140 & 0.67                        & 0.67                       & 0.65
                           \\ \midrule
\multirow{5}{*}{1}         & 4o                 & 112 & 0.81                        & 0.57                       & 0.67                          \\
                           & 3.5T               & 196 & 0.51                        & 0.74                       & 0.60                         \\
                           & L3:70b             & 160 & 0.57                        & 0.76                       & 0.66                         \\
                           & L3:8b              & 196 & 0.46                        & 0.90                       & 0.61                         \\ 
                           \noalign{\vskip -\aboverulesep}\cmidrule{2-6}\noalign{\vskip -\belowrulesep} 
                           & mean               & 166 & 0.59                        & 0.74                       & 0.64 
                           \\ \midrule
\multirow{5}{*}{2}         & 4o                 & 116 & 0.79                        & 0.57                       & 0.66                         \\
                           & 3.5T               & 185 & 0.52                        & 0.78                       & 0.63                         \\
                           & L3:70b             & 152 & 0.59                        & 0.78                       & 0.67                         \\
                           & L3:8b              & 189 & 0.48                        & 0.90                       & 0.62                         \\ 
                           \noalign{\vskip -\aboverulesep}\cmidrule{2-6}\noalign{\vskip -\belowrulesep} 
                           & mean               & 160 & 0.60                        & 0.76                       & 0.65
                           \\ \midrule
\multirow{5}{*}{3}         & gpt-40             & 120 & 0.78                        & 0.55                       & 0.64                         \\
                           & 3.5T               & 189 & 0.52                        & 0.76                       & 0.62                         \\
                           & L3:70b             & 158 & 0.58                        & 0.76                       & 0.66                         \\
                           & L3:8b              & 203 & 0.45                        & 0.88                       & 0.60                         \\ 
                           \noalign{\vskip -\aboverulesep}\cmidrule{2-6}\noalign{\vskip -\belowrulesep} 
                           & mean               & 165 & 0.58                        & 0.74                       & 0.63
                           \\ \midrule
\multirow{5}{*}{4}         & 4o                 & 116 & 0.79                        & 0.56                       & 0.66                         \\
                           & 3.5T               & 188 & 0.52                        & 0.72                       & 0.60                         \\
                           & L3:70b             & 166 & 0.56                        & 0.73                       & 0.64                         \\
                           & L3:8b              & 169 & 0.48                        & 0.83                       & 0.61                         \\ 
                           \noalign{\vskip -\aboverulesep}\cmidrule{2-6}\noalign{\vskip -\belowrulesep}
                           & mean              & 160& 0.59                        & 0.71                       & 0.63
                           \\ \bottomrule
                           
\end{tabular}

\end{wraptable}

We validated 500 manually labeled configuration dependencies of a cross-technology stack from ten real-world software projects with four state-of-the-art LLMs. Table ~\ref{tab:validation-effectiveness-all} shows the number of validation failures (\#Failures) as well as precision, recall, and F1-score for each vanilla LLM and unrefined RAG variant. Column ``RAG ID'' maps to the individual RAG systems given in Table~\ref{tab:rag-configs}. Vanilla LLMs without RAG are marked with ``w/o''.

%
%
%
%
Overall, our results show solid validation capabilities of vanilla LLMs for configuration dependencies, indicated by F1-scores ranging from 0.59 to 0.73, with \emph{GPT-4o} outperforming smaller models. When having a closer look, results are not consistent: \emph{GPT-4o} and \emph{llama3:70b} have higher precision than recall scores, which indicates that these models generally perform well in correctly identifying valid and invalid dependencies, but simultaneously miss an important portion of dependencies. By contrast, \emph{GPT-3.5-turbo} and \emph{Llama3:8b} have a high coverage in identifying valid dependencies, but produce more false positives. Although the smallest model \emph{Llama3:8b} performs best in recalling dependencies, it comes at the heavy cost of low precision. Thus, neither the vanilla LLMs nor the unrefined RAG variants provide a satisfying recall and precision. 


%

Given the unrefined RAG systems that we implemented, our results indicate that adding additional contextual information does generally not improve the validation abilities of LLMs. On contrary, in almost all cases, vanilla LLMs slightly outperform the unrefined RAG variants with the same underlying LLMs in terms of F1-score, except for \emph{LLama3:70b}. Specifically, \emph{Llama3:70b} is the only LLM that shows slightly higher F1-scores (improvements range from 0.06 to 0.09) in all RAG variants than its vanilla version. However, these improvements are only marginal and are likely not meaningful. This trend across all unrefined RAG variants is a strong indicator that even plausible context information may be unhelpful for this task. Moreover, it demonstrate that this task is quite challenging as factual relevant information seem to be not explicitly available such that the context may increase noise for an LLM. In summary, we observe that the unrefined RAG systems have a small effect on all LLMs, except \emph{GPT-4o}: they improve recall scores, meaning the models now identify more dependencies than before, but decreases precision, meaning the models now produce more false positives. For \emph{GPT-4o} there are no improvements, as all scores decrease.


%

We vary the architecture of the RAG systems in different core components to test whether these configurations affect validation effectiveness. Overall, there are only minimal variations in the validation effectiveness of LLMs. We observe only a small pattern in the fourth unrefined RAG variant, in which we decreased the number of context chunks that are provided to the LLMs. The recall of \emph{GPT-3.5-Turbo}, \emph{Llama3:8b}, and \emph{Llama3:70b} drops compared to all other RAG variants for which we provide more context chunks to the LLMs, showing that, if we provide context at all, providing more context leads to better performance. 

\subsubsection{Discussion RQ1}

The validation ability of vanilla LLMs varies significantly. While vanilla LLMs can already validate some dependencies and non-dependencies accurately, a substantial number of falsely validated dependencies remains, depending especially on the size of the LLM. 
Unrefined RAG systems have, contrary to their praise, no or mostly negative effects on the validation abilities of all LLMs, suggesting that merely adding context information does not guarantee a better outcome. This in line with related work~\cite{jain2024mitigating}, supporting the observation that an unrefined RAG system (e.g., with supoptimal retrievers or context sources) can degrade the performance of LLMs. So, simply implementing RAG systems for a concrete software engineering task is insufficient, at least, when omitting a refinement step. Poor performance with an unrefined RAG system can stem from several factors: the retriever may fail to retrieve relevant context information, the selected context sources may lack necessary information, or the prompt could be to vague, leaving room for misinterpretations. All these potentials sources of failures have to be considered when evaluating a RAG system. Therefore, a qualitative analysis of the produced failures by the RAG system is an essential part of the evaluations of RAG systems, as it sheds light not only on possible causes of failures but also helps to derive specific measures and relevant changes for effective refinements of the RAG system. 

Moreover, we have not seen substantial changes among different RAG variants. Although we cannot rule out that our changes might be too insignificant, we expect that replacing the embedding model, changing the reranker and varying the context size affect important components and should have a recognizable effect on validation performance. It seems, however, that the choice of contextual information is substantially more important than the architecture itself such that refinements should target these aspects first, instead of trying more RAG variations.




\vspace{0.2cm}
\begin{mdframed}[roundcorner=0pt]
\textbf{Answer RQ\textsubscript{1}:} 
Most vanilla LLMs exhibit better validation performance than our unrefined RAG systems with the same underlying model. Adding additional context does not automatically improve performance, particularly for more advanced LLMs. Overall, our results indicate that simply implementing an unrefined RAG system is not enough: systematically improving and refining different RAG components and context to reduce the number of failures might be key to effectively validating dependencies.
\end{mdframed}

\begin{figure}[tbh]
	\centering
	\begin{subfigure}{0.36\linewidth}
        \centering
		 \includegraphics[width=1\linewidth]{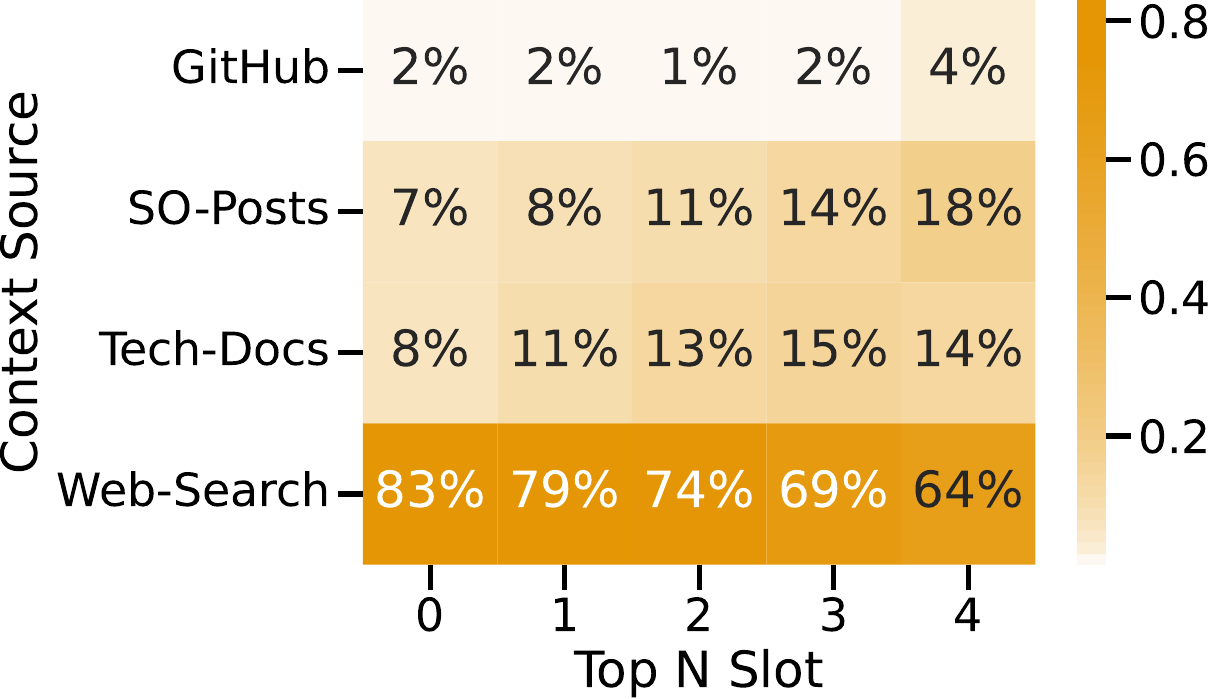}
        \caption{\footnotesize Unrefined RAG}
	       \label{fig:context}
	\end{subfigure} %
    \hfill
	\begin{subfigure}{0.27\linewidth}
        \centering
		\includegraphics[width=1\linewidth]{figures/context\_refined.pdf}
		\caption{\footnotesize Refined RAG}
	       \label{fig:contextRefined}
	\end{subfigure} %
    \hfill
	\begin{subfigure}{0.30\linewidth}
        \centering
		\includegraphics[width=1\linewidth]{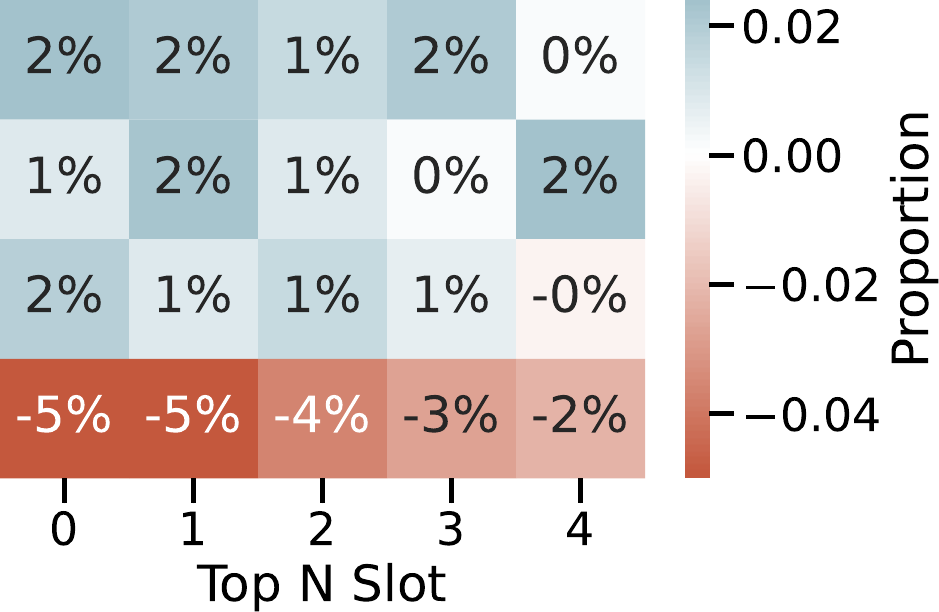}
		\caption{\footnotesize Refinement difference in p.p.}
	       \label{fig:contextRefined}
	\end{subfigure} %
 \caption{Usage of context sources per context slot for all 500 dependencies}
\end{figure}

\begin{tcolorbox}[colback=gray!5!white, colframe=gray!80!black,
                  width=\textwidth, arc=5mm, boxrule=0.5mm,
                  drop shadow=black!50!white, enhanced, 
                  title=\textbf{Reflection on the blueprint: Baseline, LLMs, and Context Sources}]
At this stage of the experiment, we can already reflect on three important aspects of the methodology: LLMs as baseline, RAG architecture, and context sources.\\
\textit{\textbf{What if we had not compared the RAG system to the vanilla LLM baseline?}}
The comparison to vanilla LLMs as baseline sets the RAG results into perspective. Looking only at different RAG variants, we could easily overestimating its impact without understanding its relative performance. Our real-world scenario prominently demonstrates that vanilla LLMs may even be superior to RAGs that include the very same LLMs. Thus, an assessment of vanilla LLMs seems essential.\\
\textit{\textbf{What if only a single LLM would have been used?}}
We see largely varying results not only among the vanilla LLMs, but also the RAG system. Also, we observe differences in LLMs depending on the RAG variant. For example, RAG variant 1 has a better recall for llama3:8b than RAG variant 4. So, even when these results are similar, we do observe interaction effects of LLMs and the RAG architecture, which can be more significant for other scenarios and, thus, should be evaluated.\\
\textit{\textbf{What about context sources?}}\\
Our evaluation has not presented results differentiated according to different context sources. In our case, as the RAG variants were inferior to the vanilla LLMs, there was no benefit in evaluating which source has which effect. In general, however, when we observe improvements, it is important to trace causes of improvements to the contextual information. To demonstrate how such an evaluation gains for insights into the RAG mechanism, we highlight in Figure~\ref{fig:context} the fraction of sources that the RAG has deemed relevant for the query and submitted to one of the five context slots. We see that most of the time, the results of the Web search are supplied to the LLM as context. Such insights represent means to evaluate the validity of the RAG system and to refine the system in an iteration.
\end{tcolorbox}

\subsection{RQ2: Validation Failures of Vanilla LLMs and unrefined RAG}
We answer the research question on which failures do occur based on a qualitative analysis. We approach this by identifying common failure categories. To this end, we manually inspect every false classification of all vanilla LLMs and RAG variant 2, as it yield the highest validation score and lowest number of false classifications among the RAG systems (see Table~\ref{tab:validation-effectiveness-all} rows ``ean''). In total, we reviewed 1\,203 validation failures from which we derived eight failure categories. Table~\ref{tab:failure-categories-statistics} depicts a summary of all categories and their distribution across the LLMs and RAG variant 2. We provide a detailed description for each category on our supplementary Web page. We see that some failures are unique to specific LLM. For instance, while \emph{GPT-4o} reliably validated dependencies in the \emph{Port Mapping}, \emph{Resource Sharing}, and \emph{Independent Technologies/Services} category, all other LLMs still struggle with these dependency types. However, there are also failure categories in which all LLMs encounter difficulties, particularly in the categories \emph{Inheritance and Overrides} and \emph{Configuration Consistency}, which show the highest number or failures. 

%
%
%
%
\begin{table}[tbh]
\caption{Failure categories of vanilla LLMs GPT-4o (4o), GPT-3.5-Turbo (3.5T), Llama3:8b (L3:8b), Llama3:70b (L3:70b), and the best found RAG variant on dependency validation.}
\label{tab:failure-categories-statistics}
\setlength{\tabcolsep}{3.0pt}
\footnotesize
\begin{tabular}{@{}lrrrrrrrr@{}}
\toprule
\multirow{2}{*}{Failure Category}         & \multicolumn{4}{c|}{Vanilla LLMs}                                          & \multicolumn{4}{c}{RAG ID 2}                                     \\ \noalign{\vskip -\aboverulesep}\cmidrule(l){2-9}\noalign{\vskip -\belowrulesep}
                                  & 4o & 3.5T & L3:8b & \multicolumn{1}{l|}{L3:70b} & 4o & 3.5T & L3:8b & L3:70b \\ \midrule
Inheritance and Overrides         & 44     & 52            & 26        & \multicolumn{1}{r|}{50}         & 49     & 42            & 30        & 36         \\
Configuration Consistency         & 25     & 46            & 54        & \multicolumn{1}{r|}{46}         & 36     & 69            & 67        & 55         \\
Resource Sharing                  & 1      & 11            & 11        & \multicolumn{1}{r|}{4}          & 6      & 15            & 15        & 9          \\
Ambiguous Options Values          & 1      & 3             & 11        & \multicolumn{1}{r|}{4}          & 2      & 8             & 12        & 7          \\
Port Mapping                      & -      & 7             & 12        & \multicolumn{1}{r|}{8}          & 7      & 10            & 8         & 7          \\
Context (Availability, Retrieval, Utilization) & 17     & 9             & 4         & \multicolumn{1}{r|}{11}         & 15     & 3             & 6         & 18         \\
Independent Technologies/Services & -      & 10            & 17        & \multicolumn{1}{r|}{10}         & -      & 17            & 10        & 6           \\
Others                            & 3      & 19            & 24        & \multicolumn{1}{r|}{21}         & 1      & 21            & 41        & 14         \\ \midrule
Total                             & 91     & 157           & 159       & \multicolumn{1}{r|}{154}        & 116    & 185           & 189       & 152         \\ \bottomrule
\end{tabular}
\end{table}

%

Across the failure categories, we developed specific RAG improvements (i.e., refinements) derived from the categories' characteristics. 
These refinements range from adding more context in the form of project-specific information, such as details about module organization and available resources over refined prompt customization to derived examples of correct and incorrect classification (i.e., few-shot prompting~\cite{brown2020language}). Providing project-specific information particularly may reduce failures in the categories \emph{Inheritance and Overrides} and \emph{Resource Sharing} as such an information is needed to perform a correct validation. Notably, we found that improving the prompt to better specify different forms of dependencies may be helpful to avoid confusions of LLMs regarding the consistency of configuration values. So, this represents a refinement that is also applicable to the vanilla LLMs and so we evaluate with this refined baseline. 

\subsubsection{Refined RAG System and Baseline}
We implemented all refinements derived from our qualitative failure analysis and applied them to the best-performing unrefined RAG system: 

\begin{compactitem}
    \item We added project-specific details by extracting relevant information about the project, including the project structure, module organization, and implementation details.  
    \item We collected exemplary dependencies alongside their correct classification and integrated them into the RAG system by adding the two most similar examples (based on cosine similarity) to the context of the LLM.
    \item  We refined the validation prompt to provide a more precise definition of dependencies. 
\end{compactitem}
Note that we also applied the last refinement to the vanilla LLMs as this is clearly a measure, which can improve the baseline and is not RAG specific. We discuss this more in the reflections on the methodology.
Next, we reran our evaluation on all prior failed dependencies using the refined vanilla LLMs and refined RAG variant 2. Not shown here, we found a substantial reduction of failures for all LLMs and the refined RAG. This rerun acts only as a sanity check since conducting an evaluation on the data from which we derived the refinements can clearly jeopardize the validity of the results. Thus, following our methodology, we assessed the final performance of both the refined vanilla LLMs and the refined RAG system on the unseen holdout set.
%
%

\begin{wraptable}[16]{r}{7cm}
\caption{Validation effectiveness of vanilla LLMs GPT-4o (4o), GPT-3.5-Turbo (3.5T), Llama3:8b (L3:8b), Llama3:70b (L3:70b)  and the refined RAG system with the same underlying LLMs on a holdout test set of 50 configuration dependencies.}
\footnotesize
\label{tab:validation-effectiveness-holdoutset}
\begin{center}
\begin{tabular}{p{0.3cm} l p{0.65cm} r r}
\toprule
RAG ID & LLM             & Pre\-cision & Recall & F1 \\ \midrule
\multirow{4}{*}{w/o} & 4o            & \multicolumn{1}{r}{0.91} & 0.65   & 0.75 \\ 
                     & 3.5T          & \multicolumn{1}{r}{0.73} & 0.61   & 0.67 \\ 
                     & L3:70b        & \multicolumn{1}{r}{0.75} & 0.58   & 0.65 \\ 
                     & L3:8b         & \multicolumn{1}{r}{0.64} & 0.89   & 0.74 \\ \midrule
\multirow{4}{*}{2}   & 4o            & \multicolumn{1}{r}{0.95} & 0.68   & 0.79 \\ 
                     & 3.5T          & \multicolumn{1}{r}{0.79} & 0.84   & 0.81 \\ 
                     & L3:70b        & \multicolumn{1}{r}{0.85} & 0.93   & 0.89 \\ 
                     & L3:8b         & \multicolumn{1}{r}{0.80} & 0.77   & 0.79 \\ \bottomrule
\end{tabular}
\end{center}
\end{wraptable}
Table~\ref{tab:validation-effectiveness-holdoutset} presents the final validation scores for all evaluated models on the holdout set. There are two notable observations: (1) The small refined vanilla LLMs especially benefit from the improved prompting. We observe a \emph{Llama3:8b} model that is now on-par with the most advanced refined model \emph{GPT-4o} and even surpasses the unrefined  \emph{GPT-4o} model for the F1 score. (2) The refined RAG systems clearly outperform the refined vanilla LLM baseline in all cases. Most notably, one of the smaller open-source models, \emph{Llama3:70b} now performs best compared to all other models with an F1 score of 0.89. Thus, both smaller open-source models combined with RAG are now a preferable option to the two large proprietary models.   

\subsubsection{Discussion of RQ2}
With our refinement, we could clearly show that a RAG system is beneficial for configuration dependency validation. Although clarifying the task improves the usage of all LLMs (especially the smaller ones), it is the additional contextual information available in a RAG system that achieves the highest scores. Although beyond the focus of this work, a further refinement concentrating on the retrieval part seems promising to reach even higher F1-scores. Especially when looking at Figure~\ref{fig:contextRefined}, we observe that the main context information is Web search, potentially leaving out other relevant information. Here, it makes sense to apply metrics, such as RAGAS~\cite{es2023ragas} to find more failure categories that are concerned with retrieval relevance. Moreover, the failure analysis is a tedious manual process, which has to be taken into account when trying to automate the development of RAG systems for certain software engineering scenarios. An artifact is that the refinements could become very specific. For our case, this could mean to jeopardize the technology-independent part. Hence, we concentrate on making all refinements agnostic to the technology at hand, but this might not be possible in any case.

\vspace{0.2cm}
\begin{mdframed}[roundcorner=0pt]
\textbf{Answer RQ\textsubscript{2}:} 
Following our blueprint, we derived simple refinements from a qualitative failure analysis and integrated them into a refined RAG variant. The refined RAG variant achieved substantial improvements compared the refined vanilla LLMs, with smaller open-source LLMs especially benefiting from these refinements, making them a preferable option when combined with RAG. Thus, knowing the failure modes of RAG allows for strong refinements and should be done in any RAG development activity.
\end{mdframed}

\begin{tcolorbox}[colback=gray!5!white, colframe=gray!80!black, 
                  drop shadow=black!50!white,
                  width=\textwidth, arc=5mm, boxrule=0.5mm,
                  enhanced,
                  title=\textbf{Reflection on the blueprint: Refinement and Assessment}]
After introducing the blueprint and demonstrating its application on a concrete software engineering task, we can reflect on the remaining aspects the of methodology: Refinement and holdout testset.\\
\textit{\textbf{Why refining the baseline is necessary?}} 
If we had not refined the baseline, our RAG system might have shown even greater improvements compared to the not refined vanilla LLMs. However, we would not be able to distinguish whether these improvements are genuinely due to enhancements in RAG-specific components or simply the result of clarifying general information in the system prompt independent of having a RAG, hence, potentially threatening the validity of our conclusions. 



\textit{\textbf{What if we had not refined the RAG system?}} Knowing in advance which contextual information actually improves the RAG performance is hard. In addition, RAG is an umbrella of different components, design decisions, and domain-specific adaptation, all of which are possible source of failures. So, unrefined RAG systems are likely to leave their potential untapped. Therefore, conducting an thorough qualitative analysis of the failures helps to incorporate the actually relevant context and select the most effective components. Furthermore, we see it important to report on the failure modes and refinements explicitly as part of a research endeavor so that others can better follow, judge, contextual, and replicate the work. Moreover, this improves also our general understand of what are best practices of RAG refinements.


\textit{\textbf{What if we had not assessed the final performance on a holdout set?}} If we had not tested the configuration validation system on a holdout set, then we would seriously threaten our external validity and generalizability. This is a rather typical scenario in a machine-learning setting for overfitting, but applies here also for the iterative refinement of RAG systems. For a practical use case, it is the testing (i.e., generalizing) error that matters. So, we need to treat our evaluation in the same way and be transparent in reporting. 

\end{tcolorbox}

\subsection{Threats to Validity}
The main threat to \emph{internal validity} is potential bias in our manual labeling of configuration dependencies. We mitigated this through rigorous review processes involving multiple authors, especially for borderline cases. We further mitigated subjectivity in our qualitative analysis of failure categories by independent coding of multiple researchers.

Regarding \emph{external validity}, our focus on software projects in the Spring Boot ecosystem may limit generalizability. Nevertheless, we selected a large range of real-world technologies that are widely used in practice. Moreover, we selected diverse, popular open-source projects as a benchmark to test on real use cases. 
%
%
The relatively small holdout test set (50 dependencies) may limit \emph{conclusion validity} and statistical power. However, the results agree with those for the large benchmark set. So, we expect no changes in data distribution. 
\begin{tcolorbox}[colback=gray!5!white, colframe=gray!80!black, 
                  width=\textwidth, arc=5mm, boxrule=0.5mm,
                  drop shadow=black!50!white,
                  enhanced,
                  title=\textbf{Reflection on the blueprint: Validity, Reliability, Objectvity}]
\textit{\textbf{Objectivity}}:
Objectivity may be threatened because the researchers who developed the blueprint are partially the researchers conducting the empirical study in the scenario. We mitigate this threat in two ways: First, we defined the methodology upfront and followed it accordingly. Second, we composed a team where one person was mainly concerned with the blueprint, one with the study on dependency validation, and one with both of them. This way, we counter a possibly too optimistic application of the blueprint.
\\
\textit{\textbf{Reliability}}:
Although it needs more studies applying our blueprint to obtain a final assessment, we have two strong indications of the blueprint's reliability: First, we derived a RAG system that challenges the state of the art in its area with only a single round of refinement --- a success on its own. Second, we reviewed upcoming papers and found similarities with the blueprint, but also methodological issues for which our blueprint can provide support.
\\
\textit{\textbf{Validity}}:
The success of the RAG system for dependency validation might also be achieved when not following our proposed research methodology. Importantly, the blueprint's goal is not to enable research that produces successful results, but sound and reliable one. That is, although if we would have reached similar results, it is due to our methodology that the results are sound and transparently reported such that one knows what measures caused them.
A threat questioning the generalizability of the blueprint is certainly the demonstration of only a single scenario. However, we made sure that this scenario has characteristics typical for the usage of RAG and is also a wide-spread use case in practice in contrast to a toy or artificial one. Additionally, our blueprint uses generic empirical concepts, which should be key aspects of any experimental evaluation and are broad enough to adapt to many scenarios. Hence, we expect the blueprint to generalize to similar scenarios without excluding the possibility of adaptations and additions in the future.
\end{tcolorbox}


\section{Conclusion}
The development of RAG systems is driven by numerous components, design decisions, and domain-specific adaptations. Without a systematic evaluation and reporting, the validity and replicability of RAG approaches are threatened. Thus, we propose a blueprint of a methodology for a sound and reliable evaluation of RAG systems. We demonstrate its applicability on a real-world software engineering task: the validation of configuration dependencies across software technologies. 

Following our blueprint, we systematically developed a RAG system that achieves the highest validation accuracy in the field of configuration dependency validation---an important research contribution on its own. At the same time, we could show how to apply the blueprint at different steps and reason on its validity. Overall, the blueprint achieved to provide sound and reliable results. It also enables a transparent reporting of different steps in the process of developing and evaluating the RAG system, thereby improving replicability and trust in the results. As a final remark, the now heavily growing number of research and studies on RAG systems requires for a more standardized and validated methodological research design. We make an important contribution toward a viable design with this paper.






\section*{Data Availability}\label{sec:data-availability}
On our supplementary website at \url{https://github.com/simisimon/dependency-validation-rag} we provide the dataset of configuration dependencies, source code for our RAG implementation, experimentation scripts, validation results, and prompt templates.

\bibliographystyle{ACM-Reference-Format}
\bibliography{references}

\end{document}